\documentclass[longbibliography,prx,showpacs,showkeys,preprintnumbers,amsmath,amssymb,superscriptaddress,twocolumn]{revtex4-1}
\usepackage{CJK} 

\usepackage{graphicx}
\usepackage{amsfonts}
\usepackage{amsmath}
\usepackage{color}
\usepackage{hyperref}
\usepackage{subfigure}

\usepackage{caption}

\usepackage{afterpage}

\def\<{\langle}
\def\>{\rangle}
\begin{document}
\begin{CJK*}{}{} 
\title{Reply to  \textquotedblleft{Comment} on \textquoteleft{Fickian} non-Gaussian diffusion in glass-forming liquids'~"}


\author{Francesco Rusciano}
\affiliation{Department of Chemical, Materials and Production Engineering, University of Naples Federico II, P.le Tecchio 80, Napoli 80125, Italy.}

\author{Raffaele Pastore}
\email{raffaele.pastore@unina.it}
\affiliation{Department of Chemical, Materials and Production Engineering, University of Naples Federico II, P.le Tecchio 80, Napoli 80125, Italy.}

\author {Francesco Greco}
\affiliation{Department of Chemical, Materials and Production Engineering, University of Naples Federico II, P.le Tecchio 80, Napoli 80125, Italy.}

\begin{abstract}
In ~\cite{berthier2022supercooled}, Berthier et al. questioned the findings of our letter~\cite{rusciano2022fickian}, concerning the existence and the features of Fickian non-Gaussian diffusion in glass-forming liquids.
Here we demonstrate that their arguments are either wrong, or not meaningful to our scope. Thus, we fully confirm the validity and novelty of our results. 
\end{abstract}


\maketitle
\end{CJK*} 

Fickian non-Gaussian Diffusion (FnGD), also known as Brownian non-Gaussian (BnG) diffusion, is commonly associated to some heterogeneity of the environment where tracers diffuse. In glassy materials, dynamic heterogeneity (DH) is a crucial feature, thoroughly investigated in its many facets. Such a good characterization of heterogeneity is exactly our starting point to search for FnGD in glass-forming liquids.

In our article~\cite{rusciano2022fickian}, we showed that the long-time diffusion of two glass-forming systems displays an increasingly marked FnGD regime on approaching the glass transition. 
The main results of ~\cite{rusciano2022fickian} are: i) the identification of two characteristic timescales for the onset of Fickianity (i.e., linear time-dependence of the MSD) and the recovery of Gaussianity, $\tau_F$ and $\tau_G$, respectively, ii) the existence of a power-law relation $\tau_G \propto \tau_F^{\gamma}$ (with $\gamma>1$), and iii) the collapse of the decay lengths $l$ of the Van Hove function exponential tails over a power-law master-curve $l(t) \propto t^{0.33}$ in the FnGD regime.

In their comment~\cite{berthier2022supercooled}, Berthier, Flenner and Szamel (BFS) questioned the novelty of our findings, 
also stating that the identified time- and length-scales are ill-defined. In fact, they go so far as \textquotedblleft{}\textit{to dispute the conclusion that glass-formers display Fickian non-Gaussian behaviour}".

After having considered their comment,  we remain fully convinced that it is the very first time that our findings i), ii) and iii) are reported, not only for glassy liquids, but also in the broad context of FnGD.
We also deem that the technical concerns raised by BFS are not well-grounded and, therefore, we do reaffirm that long-time dynamics in glass-formers is definitely Fickian and non-Gaussian. 
In the following, we support our just given conclusions. 
\\

In questioning that FnGD is actually present in glass-forming liquids, BFS notice that apparent Fickianity in these systems is preceded by a subdiffusive regime,
with an ensuing \textquotedblleft{}\textit{algebraic approach}" to linearity in the MSD.
However, it is worth reminding that Fickian dynamics in actual systems is \textit{always} preceded by some anomalous diffusion, not only in glass-forming liquids
\footnote{In this perspective, glass-forming liquids are qualitatively similar to any other system.
Differences are merely quantitative:
in some systems, for instance, pre-Fickian  regimes are limited to very short times and migth be hardly detectable in experiments. In liquids approaching the glass transition, conversely, subdiffusion becomes long-lasting and easily detectable.
Interestingly, a similar pre-Fickian subdiffusive regime is also observed in other systems displaying FnGD at longer times~\cite{wang2009anomalous,song2019transport,pastore2021rapid}, including the actin networks investigated by Granick group in their seminal paper on FnGD~\cite{wang2009anomalous}.}.
BFS further state that:
\textquotedblleft{}\textit{strictly speaking, self-diffusion is never Fickian at any time $ <\infty$}" and \textquotedblleft{}\textit{is never Gaussian at any time $<\infty$}".
This asymptotic argument is, in fact, \textquotedblleft{}\textit{only a mathematical conclusion, i.e. a conclusion from an infinitely long trajectory. But \textquoteleft{infinitely} long' has no physical meaning}"~\cite{ma1985statistical}.
Indeed, the statements by BFS would unavoidably lead to the denying of the very existence of FnGD. 
Needless to say, in spite of the presence of pre-Fickian and pre-Gaussian regimes, Fickianity and Gaussianity are observed to be fully recovered at  finite times in many actual systems (both from experiments and simulations), within obvious uncertainties inherent to any measurements.
Thus, it is fully legitimate and meaningful to measure those two finite times $\tau_F$ and $\tau_G$, which we in fact find to be distinct from one another.
It is worth noticing that BFS themselves used the phrasing  \textquotedblleft{}\textit{time/length-scale of the onset of Fickian diffusion in supercooled liquids}" to mark finite time/length-scales in some of their papers~\cite{berthier2004length,berthier2004time,szamel2006time}.

Concerning the time $\tau_F=\frac{\sigma^2}{2dD}$, with $d$ being the dimensionality, we are not introducing a novel timescale; rather, we demonstrate that, in the investigated systems, $\tau_F$ marks (within obvious uncertainties) the onset of linearity in the MSD, while Gaussianity is still far from being reached.

For what matters the time $\tau_G$ for the recovering of Gaussianity, we reject the claim by BFS that it is an ill-defined quantity.
Indeed, $\tau_G$ is defined well within an observed power-law regime in the late decay of the non-Gaussian parameter $\alpha_2(t)$. This regime is always attained well after $\tau_F$, which falls instead close to the maximum of $\alpha_2(t)$.
Thus,  it is $\tau_G > \tau_F$ systematically, regardless of the adopted threshold value for $\alpha_2$. 
In addition, just because long-time $\alpha_2(t)$ data for different temperature/area fraction collapse onto a unique master-curve (being it a power-law is not essential), the temperature/area fraction dependence of $\tau_G$ is independent on the adopted threshold. As a matter of fact, our approach draws on a  \textquotedblleft{}time-temperature/concentration" superposition 
with its shift factors, in analogy with other characteristic scales defined in a variety of systems \cite{doi1988theory, charbonneau2014hopping,dyre2000universality}.
Overall, the above mentioned features make the characteristic time $\tau_G$ both well-defined and robust.
\\

Having clarified that the characteristic timescales studied in~\cite{rusciano2022fickian} are well-defined, we now show that the main criticisms raised by BFS are due to a quantitative error committed in~\cite{berthier2022supercooled},
which leads to misunderstand our results and their novelty.
Indeed, BFS write \textquotedblleft{}\textit{Linear behaviour of the MSD may be visually detected in log-log representations after a time $\tau_D$ which
is indeed smaller than $\tau_{\alpha}$}".
(Notice that BFS define $\tau_D=\sigma^2/D$, hence $\tau_D=2d\tau_{F}$; $\tau_{\alpha}$ is the traditional structural relaxation time).
This statement is wrong: indeed, in the model systems investigated in our article~\cite{rusciano2022fickian}, $\tau_F$, and even more so $\tau_D$, is always significantly larger than $\tau_{\alpha}$.

Precisely, it is $\tau_F/\tau_{\alpha}\simeq 10$ for the experimental system over the entire area fraction range,
while  $\tau_F/\tau_{\alpha}$ varies from $\simeq 12$ to $\simeq 2$ on lowering the temperature in simulations.
Even in the most paradigmatic model of glass-forming liquids, i.e. the 3D Kob-Andersen Lennard Jones mixture\cite{kob1994scaling}, and at the lowest temperature achieved in~\cite{coslovich2018dynamic} (using advanced numerical techiques)  $\tau_F$ is larger than $\tau_{\alpha}$\footnote{Precisely, as it can be appreciated from Ref.s~\cite{coslovich2018dynamic,porpora2022comparing},  $\tau_F/\tau_{\alpha} \simeq 12$ (hence  $\tau_D/\tau_{\alpha} \simeq 72$)  at around the onset temperature $T=0.7$, and $\tau_F/\tau_{\alpha} \simeq 2$ (hence  $\tau_D/\tau_{\alpha} \simeq 12$) at the lowest temperature $T=0.39$. }.
Thus, while in some glass-formers and in some control-parameter ranges $\tau_{\alpha}$ may increase faster than $\tau_D$, as a matter of fact, BFS's claim that $\tau_{\alpha}>\tau_D$ is generally wrong. 

This mistake by BFS leads to faults in their successive criticisms about the novelty of our results with respect to previous works~\cite{berthier2004length,chaudhuri2007universal,chaudhuri2008random,hedges2007decoupling,berthier2004time,szamel2006time}.
It is worth noticing that, in most of those works, $\tau_{\alpha}$ is supposed to control the slowest investigated dynamics.
By contrast, this time is not relevant in our letter~\cite{rusciano2022fickian}, which in fact focuses on the dynamics on much longer time scales. Precisely, $\tau_\alpha$ is well below the lower boundary $\tau_F$ of the time-range considered in our work, i.e. $[\tau_F,~\tau_G]$. 
Notice that this point is already fully highlighted in our article, where we write  that $\tau_\alpha$  \textquotedblleft{}\textit{typically falls either within the  subdiffusive regime or around the subdiffusive-Fickian crossover}"~\cite{rusciano2022fickian}.

Moreover, we do remark that our analysis deal with quantities specifically targeted to spot out FnGD, whereas those papers~\cite{berthier2004length,chaudhuri2007universal,chaudhuri2008random,hedges2007decoupling,berthier2004time,szamel2006time}, appeared quite earlier than the discovery of FnGD~\cite{wang2009anomalous}, focus of course on different quantities and different scopes.   
More specifically, BFS are wrong when they state that  we are \textquotedblleft{}\textit{rediscovering}" a result of~\cite{szamel2006time}, in finding $\tau_G$ to increase faster than $\tau_F$.
In Ref.~\cite{szamel2006time}, in fact,  Szamel and Flenner studied  $\tau_{\alpha}$ and a so-called \textquotedblleft{}Fickian time", which is in fact  defined from the restoring of Gaussianity, with no reference to the restoring of linearity in the MSD.
Indeed, Szamel and Flenner wrote in their article~\cite{szamel2006time}: \textit{The probability distribution
$P(log\Delta r;t)$ is a convenient indicator of Fickian
diffusion, because if particles move via Fickian diffusion,
then the self-part of the van Hove function is Gaussian"}
As a matter of fact,  with \textit{Fickian diffusion} they meant \textit{Fickian and Gaussian diffusion}
This latter naming mirrors the belief in a coincidence of Fickian and Gaussian diffusion, as it was widespread before the discovery of FnGD~\cite{wang2009anomalous}. 
Interestingly, in~\cite{szamel2006time} it is also hinted that the onset time of Gaussian diffusion \textquotedblleft{}\textit{occurs well within the regime of
apparent linear time dependence of the mean-squared displacement}".
Nevertheless, no quantitative analysis of this issue (primarily, the time needed to recover the linearity in the MSD) is provided in Ref.~\cite{szamel2006time}, which is instead exactly what we did in our article.
At the same time, the above quotation fully supports our motivations, namely the quantitative search for Fickian non-Gaussian regime in glass-forming liquids. 

Another wrong claim included in~\cite{berthier2022supercooled} is that the lengths defined in Ref.s~\cite{berthier2004length,berthier2004time}
are equivalent to the Gaussian length-scale $\xi_G$ defined in our letter. 
Indeed, our $\xi_G$ and the length defined in~\cite{berthier2004length} are the the root MSDs evaluated at $t=\tau_G$ and at $t=\tau_{\alpha}$, respectively. Since the two timescales are different ($\tau_G \gg \tau_{\alpha}$, including a generally different temperature dependence), the two lengths are also intrinsically different.
On the other hand, the length defined in~\cite{berthier2004time} is related to the characteristic size of dynamical heterogeneity (obtained from a multi-point correlation function), and, accordingly, it is also intrinsically different from our $\xi_G$. 

Concerning the exponential decay length $l(t)$ of the Van Hove function, BFS question that our measurements are 
meaningful. We notice that, in doing this criticism, they explicitly refer to what occurs \textquotedblleft{}\textit{at }$t \leq \tau_{\alpha}$" and to some of their works~\cite{chaudhuri2007universal,chaudhuri2008random} that focused on exponential tails in this time range. As already clarified, this time range is not considered at all in our work and, therefore, the criticism is poorly relevant. 
More importantly, we remark that our measurement of $l(t)$ is a very robust one. Indeed, we were careful in performing our fitting in a time-range $\tau_F\leq t < \tau_G$ (i.e. ending well before $\tau_G$) where exponential tails are well developed and far from reverting to a Gaussian distribution.
\\

Along the comment~\cite{berthier2022supercooled}, there are also a number of other minor criticisms consisting in a few sentences that are not supported by any quantitative arguments, as we discuss in the following.

Concerning the criticism that our results (obtained for two-dimensional systems) may be affected by Mermin-Wagner fluctuations, we notice that such fluctuations are found to have an impact on the dynamics at relatively short times, especially consisting in a weakening of the \textit{cage}-effect, if compared to three-dimensional systems~\cite{flenner2015fundamental}. Thus,  this criticism is irrelevant for our results, which concern with very long-time dynamics, when \textquotedblleft{}\textit{on average, particles have performed many jumps}", having definitely escaped their original cages (as explicitly clarified in~\cite{rusciano2022fickian}).

 BFS also write that self-diffusion in glass forminng liquids \textquotedblleft{}\textit{is very different from several other
Fickian non Gaussian materials where exogenous disorder modifies self-diffusion in an interesting manner}~\cite{chechkin2017brownian}".
However, it is worth remarking that the FnGD model of~\cite{chechkin2017brownian} (as well as the original Diffusing Diffusivity model~\cite{chubynsky2014diffusing}) do not make any distinction between \textquotedblleft{}exogenous" and \textquotedblleft{}endogenous" disorder, as instead suggested in~\cite{berthier2022supercooled}.
As a matter of fact, the interpretation by BFS seems to be in stark contradiction with recent papers~\cite{miotto2021length,sposini2022detecting} by some of the authors of~\cite{chechkin2017brownian}. For instance, verbatim from~\cite{sposini2022detecting}:  \textquotedblleft{}\textit{Given that the 
emergence of dynamic heterogeneities in glass-forming
liquids upon cooling is a well established phenomenon~\cite{kob1997dynamical,kegel2000direct}, 
the connection between BnG diffusion, exponential-tailed 
distributions and glass-forming liquids helps in strengthening the standard interpretation of BnG diffusion as due to heterogeneity in the system. In addition, such connection opens new directions of study since the overall picture of this phenomenology is still far from being completely understood}".

In ~\cite{berthier2022supercooled}, it is also stated that \textquotedblleft{}\textit{Known results~\cite{berthier2004length,chaudhuri2007universal,chaudhuri2008random,hedges2007decoupling,berthier2004time,szamel2006time,helfferich2014continuous} paint a  picture that is inconsistent with several findings}" of our work~\cite{rusciano2022fickian}, but those inconsistecies are not declared at all. 
\\

Rather, we believe to have complemented the \textquotedblleft{}known picture" of DHs in glass-forming liquids with novel results, at a single-particle level, on a previously poorly investigated regime. At the same time, we believe to have provided interesting hints to the the much broader and still puzzling issue of FnGD. 
In particular, we hope to have clarified how glass-forming liquids are a privileged stage for advancing the understanding of FnGD.

\bibliography{apssamp.bib}

\end{document}